\documentclass[prl,twocolumn,superscriptaddress]{revtex4}
\usepackage{epsfig}
\usepackage{times}
\usepackage{amsmath}
\usepackage{amsfonts}
\usepackage{amssymb}
\usepackage[usenames]{color}
\usepackage{graphicx}

\newcommand{\beq}{\begin{equation}}
\newcommand{\eeq}{\end{equation}}
\newcommand{\beqa}{\begin{eqnarray}}
\newcommand{\eeqa}{\end{eqnarray}}

\begin{document}
\title{Enhanced energy relaxation  process of quantum memory 
coupled with a superconducting qubit}
\author{Yuichiro Matsuzaki}
 \affiliation{
  NTT Basic Research Laboratories, NTT Corporation, 
  Kanagawa, 243-0198, Japan
 }
\author{Hayato Nakano}
  \affiliation{
  NTT Basic Research Laboratories, NTT Corporation, 
  Kanagawa, 243-0198, Japan
  }

\begin{abstract}
 For quantum information processing, each physical system has
 a different advantage
 as regards implementation and so hybrid systems
 that benefit from the advantage of several systems
 would provide a
 promising approach.  One common hybrid approach
  involves combining a
 superconducting qubit as a controllable qubit and
 another quantum system with a long coherence time as
 a memory qubit.
 The use of a superconducting qubit
  gives us excellent controllability of the quantum states and the
 memory qubit is capable of storing information for a long time.
  It has been believed that selective coupling can be
 realized between a superconducting qubit and a memory qubit by tuning
 the energy splitting between them.
 However, we have shown that this \textcolor{black}{detuning} approach has a fundamental drawback
  as regards energy leakage from the memory qubit.
 \textcolor{black}{Even if the superconducting qubit is effectively
 separated by reasonable detuning, a non-negligible incoherent energy relaxation
 in the memory qubit occurs via residual
 weak coupling when the superconducting qubit is affected by severe
 dephasing. }
 \textcolor{black}{
 This energy transport from the memory qubit to the control qubit can be interpreted as the appearance
 of the anti
 quantum Zeno effect induced by the fluctuation in the superconducting qubit.}
 We also discuss possible ways to avoid  this energy
 relaxation process, which is feasible with existing technology. 
\end{abstract}

\maketitle

\section{I. introduction}
 A superconducting qubit
provides us with excellent controllability of the system for quantum
information processing.
 Coherent
 manipulations of superconducting qubits
 have already been demonstrated experimentally, and actually
  it is possible to perform a single qubit rotation
 within a few nano seconds by using a
resonant microwave field \cite{kutsuzawa-rotation}.
 Also, a high fidelity single qubit measurement has already
 been achieved with existing technology
 \cite{ClarkeWilhelm01a}.
  Specifically, a method using a Josephson
 Bifurcation Amplifier (JBA)
 \cite{SiddiqiVijayPierreWilsonMetcalfeRigettiFrunzioDevoret01a,LupacuSaitoPicotGrootHarmansMooij01a}
 has been used experimentally to perform a non-destructive measurement
 on the superconducting qubit.
 However, the coherence time of the superconducting qubit is usually
 relatively short where the typical dephasing time is
 of the order of a microsecond at the optimal point and becomes tens of nanoseconds far from
 the optimal point
 \cite{KakuyanagiMenoSaitoNakanoSembaTakayanagiDeppeShnirman01a,YoshiharaHarrabiNiskanenNakamura01a,
 astafiev2004quantum}.

 Recently, to overcome the problem of the short coherence time,
 a hybrid
 approach has been suggested that one can use
  another physical system as a quantum
 memory.
 \textcolor{black}{One promising system for quantum memory is an atomic
 ensemble of electronic spins such as P-doped Si and nitrogen atoms
 in fullerene cages C${}_{60}$  where the spin ensemble is
 coupled with the superconducting qubit through a microwave cavity \cite{tyryshkin2003electron,
 morton2008solid,morton2007environmental}.
 Magnetic coupling between a
 superconducting flux qubit and a spin ensemble such as nitrogen-vacancy centers in
 diamond can also provide such a hybrid system \cite{marcos2010coupling, zhu2011coherent}.
 Spin ensemble qubits typically have
 a long coherence time of, for example, tens of milliseconds, which is a million
 times longer than that of a superconducting qubit
 \cite{tyryshkin2003electron, morton2008solid,morton2007environmental,
 marcos2010coupling}. Moreover, for the electron spins bound to donors
 in silicon, coherence times can be as long as several seconds
 \cite{tyryshkin2011electron}.
}
   It is known that a superconducting qubit could potentially provide a
   memory qubit if the lifetime could be
   increased. The control and the measurement setup used for the
   superconducting qubit, however, induces decoherence, and so there is a trade off
   relationship between efficient control and a long coherence time \cite{van2003engineering}.
   This means that, by sacrificing controllability, it would be
   possible to have a 
   much longer coherence time for a memory superconducting qubit \cite{van2003engineering,Bylander2011dynamicalnoise}.
   By combining a superconducting qubit with excellent controllability 
   and another superconducting qubit with a long coherence time, we can
   construct a hybrid system to take advantage of both
   characteristics.
   \textcolor{black}{
   For example, a recent experiment coupling two superconducting phase qubits with a
 resonant cavity \cite{altomare2010tripartite} showed the possibility of
 utilizing one of the qubits for control and the other for memory
 where two phase qubits are entangled through the common quantum bus,
 namely the resonant cavity \cite{blais2007quantum}.}
   Another example is a hybrid system consisting of a superconducting qubit and a 
microwave cavity. Strong coupling has been realized between the superconducting qubit and
microwave cavity \cite{wallraff2004strong,
 houck2007generating}, which shows a possible application of
 a high Q cavity as a quantum memory
 for storing the information.

 However, in this paper, we \textcolor{black}{point out quantitatively} that such hybrid systems composed of
 a superconducting qubit and a memory qubit
 \textcolor{black}{could} have a potential error caused by unwanted energy leakage from the memory
 qubit.
 \textcolor{black}{When we} transfer the quantum information to the memory,
 we have to
 tune the energy of the superconducting qubit on resonance with the memory
 qubit, and then it becomes possible to swap the information from the
 controllable qubit to the memory qubit. 
 \textcolor{black}{Subsequently, by changing external magnetic field, we can detune the
 energy of the superconducting qubit to decouple from the memory qubit.}
 Importantly, the superconducting qubit is usually affected by
 severe dephasing
 \cite{KakuyanagiMenoSaitoNakanoSembaTakayanagiDeppeShnirman01a,YoshiharaHarrabiNiskanenNakamura01a},
 and this induces an incoherent energy \textcolor{black}{leakage}
 from the memory qubit to the superconducting qubit during
 information storage.
 \textcolor{black}{
 This energy relaxation caused by dephasing violates the energy
 conservation}, and this phenomenon can be understood as an occurrence of 
 anti-Zeno effect for quantum transport
 \cite{plenio2008dephasing,rebentrost2009environment,caruso2009highly,fujii2010anti}.
 With reasonable
 experimental parameters, we \textcolor{black}{evaluate} the actual lifetime of the
 memory qubit, and this turns out to be
 much shorter than the previously expected lifetime
 of the memory qubits
 \cite{tyryshkin2003electron, morton2008solid,morton2007environmental, marcos2010coupling}.
  We will suggest possible ways to avoid such an energy
 relaxation process, which is feasible with
 current technology.

 \textcolor{black}{The remainder of this paper is organized as follows. In
 Sec II, we review the concept of quantum Zeno and anti Zeno effects. 
 Sec III presents the details of our calculations to show how unwanted
 relaxation occurs in the memory qubit due to the instability of the
 detuned but weakly coupled superconducting qubit. In Sec IV, we suggest
 some ways to avoid
 such relaxation by using the idea of the decoherence free subspace. 
 Sec V concludes our discussion.}
            \begin{figure}[h]
       \begin{center}
        \includegraphics[width=8.75cm]{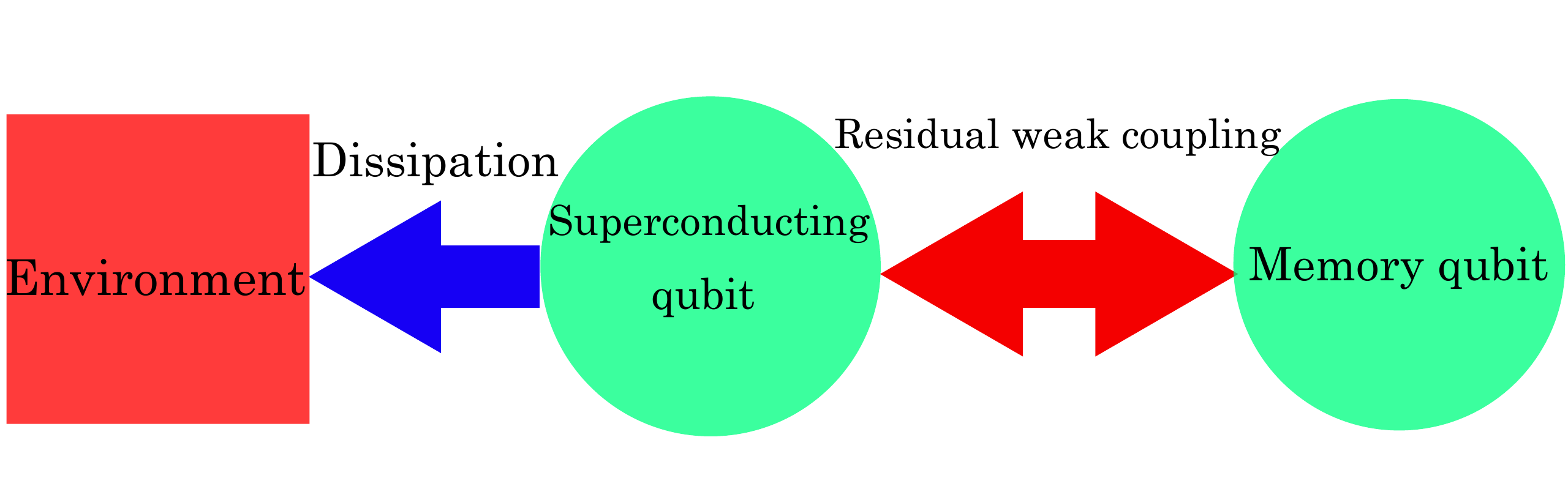} 
        \caption{Schematic of the enhanced relaxation of the memory qubit via
        imperfect decoupling from a superconducting qubit.
        We assume that the superconducting qubit is affected by decoherence
        while there is no direct coupling between the environment and the
        memory qubit.
        Since we can
        tune the energy of the superconducting qubit, it is possible to make
        the energy of the superconducting qubit on resonance with the energy
        of the memory qubit to transfer the information. It is also possible
        to tune the energy of the superconducting qubit far from the
        resonance when \textcolor{black}{keeping the stored} information. However, this type of
        selective coupling has a significant drawback \textcolor{black}{during the information
        storage in the memory qubit, as explained in the main text}.
        }\label{antizeno-schema}
       \end{center}
       \end{figure}
\section{II. Quantum Zeno and anti Zeno effects}
\textcolor{black}{Quantum Zeno and anti Zeno effects are fascinating
phenomena predicted by quantum mechanics \cite{BECG01a,Kofman01a,Cook01a}.
Let us summarize the quantum Zeno and anti Zeno effects.
When an unstable excited state decays to a ground state, we can
define the survival probability $P_{\text{e}}(t)$ as the population remaining in the excited
state at time $t$.
If this survival probability exhibits a quadratic behavior in the initial
stage of the decay such as $P_{\text{e}}(t)\simeq 1-\Gamma ^2t^2$ for $\Gamma
t\ll 1$, it is possible to confine the state to the excited level via 
frequent projective measurements.
When we perform $N$ projective measurements with a time interval $\tau
=\frac{t}{N}$
to determine whether or not the state
is still in the excited state, the probability of projecting the state in
the excited level for all $N$ measurements is calculated as $P(t,N)\simeq
(1-\Gamma ^2 \tau ^2)^N\simeq 1-\Gamma ^2 \frac{t^2}{N}$.
This success
probability approaches unity as the number of
measurements increases. So the system is frozen and decay can be completely
suppressed, which is called the quantum Zeno effect.
On the other hand, if the survival probability exhibits an
exponential decay such as $P_{\text{e}}(t)=e^{-\Gamma t}$, the probability of confining the state to the excited
level by performing $N$ measurements is calculated as
$P(t,N)=(e^{-\Gamma \tau })^N=e^{-\Gamma t}$. Here,
the projective measurements do not change the success probability,
which means that 
we cannot observe quantum Zeno effect for such an exponential decay
system.
 Interestingly, it is known that unstable systems show quadratic
decay initially and exponential decay later
\cite{NakazatoNamikiPascazio01a}.
The temporal scale used to denote the crossover from quadratic to exponential
decay is known as the jump time \cite{Schulman01a}.
Therefore, to observe the quantum Zeno effect, it is necessary to perform
projective measurements on a time scale shorter than the jump time.
Moreover, it was predicted that, when projective measurements are performed
on a time scale comparable to the jump time, the decay is effectively
accelerated, and this is called the anti quantum Zeno effect \cite{Kofman01a, kaulakys1997quantum}.
Recently, it was also predicted that the
 anti quantum Zeno
effect can be induced when we perform false measurements
 \cite{koshino2004quantum}, namely,
 the decay dynamics of the unstable state can be enhanced by frequent measurements with
an erroneous apparatus
 where
the energy band of the measurement apparatus is significantly different from
the energy of the signal emitted from the unstable state.
}
\section{III. Enhanced energy relaxation process}
 Let us study the unexpected relaxation of the memory qubit
 \textcolor{black}{in a hybrid system}
 shown in Fig. \ref{antizeno-schema} quantitatively.
 Although such relaxation behavior has been studied by
 \cite{plenio2008dephasing,rebentrost2009environment,caruso2009highly,fujii2010anti}
 in an anti Zeno context for quantum energy transport,
 we introduce a
 simpler solvable model and we derive an analytical form of the energy
 relaxation time of the memory qubit.
 To describe the coupling between the superconducting
 qubit and the memory qubit, we use the following Hamiltonian called the Jaynes-Cummings model
or the Tavis-Cummings model
 \begin{eqnarray}
  H=\frac{\omega _{\text{sc}}}{2} \hat{\sigma }_z ^{\text{(sc)}}
   +\frac{\omega
   _{\text{\text{m}}}}{2}  \hat{\sigma }_z^{\text{(m)}}
   +g(\hat{\sigma }_+^{(\text{sc})}\hat{\sigma }_-^{(\text{m})}+
   \hat{\sigma }_-^{(\text{sc})}\hat{\sigma }_+^{(\text{m})} )
   \label{hamiltonian}
 \end{eqnarray}
 where $\omega ^{(\text{sc})}$ ($\omega ^{(\text{m})}$)
 denotes the energy of the superconducting qubit (memory qubit) and $g$
 denotes the coupling strength of the interaction.
\textcolor{black}{Note that, although we refer to a superconducting qubit as a
 control system coupled with the memory in this paper, the analysis here can be applied to
 any system
 as long as the interaction with the memory device
 is described by the Jaynes-Cummings model
or the Tavis-Cummings model.
These models are of fundamental importance not only for the present setup
 but also for many variations: coupling between
 superconducting qubits \cite{neeley2010generationetal,steffen2006measurement}: a
 superconducting qubit interacting with a microwave cavity
 \cite{altomare2010tripartite,blais2007quantum}, a superconducting resonator coupled with
 a spin ensemble \cite{kubo2010strong,schuster2010high, abe2011electron}, or a
 superconducting flux qubit magnetically coupled with nitrogen-vacancy centers in
 diamond \cite{marcos2010coupling}.}
Since we can change the energy of the superconducting qubit, it is
possible to detune the energy between qubits when keeping the information
stored in the memory.
In this paper, $\Delta =\omega _{\text{sc}} -\omega _{\text{m}}$ denotes detuning during such a
storage.
We choose the initial state as $|0\rangle _{\text{sc}}|1\rangle _{\text{m}}$ to
represent the storage of the excitation in the memory qubit. Note that,
since the Hamiltonian conserves the total number of the excitation, the
bases taken into account are $|0\rangle
_{\text{sc}}|1\rangle _{\text{m}}$ and $|1\rangle _{\text{sc}}|0\rangle _{\text{m}}$,
as long as we consider only the dephasing of the superconducting qubit as
the decoherence source. \textcolor{black}{In other words, the state of the
coupled system is always in the Hilbert subspace spanned by these two bases.}
Also, it is worth mentioning that, throughout this paper, we assume that
the memory
qubit is coupled only with the superconducting qubit and
has no direct interaction with the environment. The assumption here is
valid as long as the lifetime of the memory qubit is much longer than
that of a superconducting qubit. \textcolor{black}{This is actually true for
typical memory qubits because
 the coherence time of memory qubits can be tens of milliseconds,
 which is a million
 times longer than that of superconducting qubits
 \cite{tyryshkin2003electron, morton2008solid,morton2007environmental,
 marcos2010coupling}. }
To obtain an analytical solution of the dynamics under the effect of the dephasing, we adopt a simple model where the
system is affected by the unitary operation and the decoherence
alternatively, so that we can obtain a recursion equation
as $ \rho _{(n+1)\tau} =\hat{\mathcal{E}}(e^{-iH \tau }\rho _{n \tau
  }e^{iH\tau })$. 
Here, $\rho _{n\tau }$ denotes the density matrix of the system at time
$n\tau $,
$\hat{\mathcal{E}}$ denotes the dephasing process, and $\tau $ denotes a
period during the unitary operation.
\textcolor{black}{ In the limit for
a small $\tau $, this simplification can be justified by the Trotter
expansion \cite{trotter1959}.
Moreover, the effect of dephasing can be considered 
a process of measurements without postselection, which we refer to as a
``non-selective measurement''.
For example, if a pure state $\alpha |0\rangle +\beta |1\rangle $
decoheres due to the dephasing, we finally obtain a mixed state $|\alpha
|^2|0\rangle \langle 0|+|\beta |^2|1\rangle \langle 1|$, which is the
same state as that obtained after performing a projective measurement
with respect to \textcolor{black}{$\hat{\sigma }_z=|0\rangle \langle 0|-|1\rangle \langle 1|$} on the
pure state and
discarding the measurement results \cite{vNbook32springer, koshino2005quantumshimizu}.
Therefore, our model can be interpreted as one where the environment ``sees'' the system frequently to degrade the quantum
  coherence, which provides us with an intuitive connection between our
  calculation and the quantum Zeno effect. When}
 the time $\tau $ is comparable to the dephasing time
$T^{(\text{sc})}_2$ of the superconducting qubit such as $\tau =\alpha T^{(\text{sc})}_2$ where
$\alpha $ is a fitting parameter \textcolor{black}{of the order of unity},
the off-diagonal terms of the density matrix become small.
So we consider the superoperator
$\mathcal{\hat{E}}$ to be a non-selective measurement process for removing out
the off-diagonal terms as follows;
\begin{eqnarray}
 \mathcal{\hat{E}}(\rho _{n\tau })&\simeq &(|0\rangle _{\text{sc}}\langle
  0|\otimes \hat{\openone} _{m})\rho _{n\tau }(|0\rangle _{\text{sc}}\langle
  0|\otimes \hat{\openone} _{m})\nonumber \\
  &+&
  (|1\rangle _{\text{sc}}\langle
  1|\otimes \hat{\openone} _{m})\rho _{n\tau }(|1\rangle _{\text{sc}}\langle
  1|\otimes \hat{\openone} _{m}) \nonumber \\
 &=&\mathcal{\hat{P}}^{\text{(sc)}}_0\rho _{n\tau }\mathcal{\hat{P}}^{\text{(sc)}}_0 +
  \mathcal{\hat{P}}^{\text{(sc)}}_1\rho _{n\tau }\mathcal{\hat{P}}^{\text{(sc)}}_1 
\end{eqnarray}
where $\mathcal{\hat{P}}^{\text{(sc)}}_0$
($\mathcal{\hat{P}}^{\text{(sc)}}_1$) is the projection operator to a
state $|0\rangle _{\text{sc}}$ ($|1\rangle _{\text{sc}}$).
Under this approximation, the mixed state after performing this
superoperator $\mathcal{\hat{E}}$
should be described as $\rho _{n\tau }=p_{a, n\tau } |10\rangle
_{sc,m}\langle 10| + p_{b,n\tau } |01\rangle
_{sc,m}\langle 01|$ where $p_{a, n\tau }$ and $p_{b, n\tau }$ denote the
population of each state.
So we obtain the recursion equations as follows:
\begin{eqnarray}
 p_{a,(n+1)\tau }&=&\frac{1}{4g^2+\Delta  ^2}(2g^2+\Delta  ^2 p_{a,n\tau
  }\nonumber \\
  &+&2g^2(p_{a,n\tau
  }-p_{b,n\tau }) \cos t\sqrt{4g^2+\Delta  ^2})
  \nonumber \\
  p_{b,(n+1)\tau }&=&\frac{1}{4g^2+\Delta  ^2}(2g^2+\Delta  ^2 p_{b,n\tau
  }\nonumber \\
  &-&2g^2(p_{a,n\tau
  }-p_{b,n\tau }) \cos t\sqrt{4g^2+\Delta  ^2}).
  \nonumber
\end{eqnarray}
By solving these equations with the initial condition of $p_{a,0}=0$ and
$p_{b,0}=1$, we obtain
\begin{eqnarray}
 p_{a,n}&\simeq &\frac{1}{2}(1-(\frac{\Delta  ^2}{4g
^2 +\Delta 
  ^2})^n)\nonumber \\
 p_{b,n}&\simeq &\frac{1}{2}(1+(\frac{\Delta  ^2}{4g ^2
  +\Delta  ^2})^n)
\end{eqnarray}
  where we use a rotating wave approximation such that $ \cos \sqrt{4g^2
  +\Delta  ^2} t$ should vanish due to the high frequency oscillation.
  We define the effective relaxation time induced by this anti Zeno
  effect as the time at which the population of the excitation of the memory
  becomes $\frac{p_{b,0}-p_{b,\infty }}{2}$.
  So we can calculate this effective relaxation time
  \textcolor{black}{of the memory qubit}
  as
  \begin{eqnarray}
   \tilde{T}^{\text{(m)}}_1&=&\frac{\alpha \log 2}{\log
  (1+\frac{4g^2}{\Delta 
  ^2})} T^{(\text{sc})}_2\nonumber \\
  &\simeq &\alpha \log 2 \cdot \frac{\Delta ^2}{4g ^2}T^{(\text{sc})}_2 .\label{effective-relaxation-time}
  \end{eqnarray}
   \textcolor{black}{Here,
  we assumed $\frac{g}{\Delta }\ll 1$, namely, the coupling is much smaller
  than the detuning, which is appropriate for actual
  experiments in order to decouple the system.}
  Although the energy relaxation \textcolor{black}{might be considered} to be exponentially small for a
  large detuning,
  the effective relaxation time of the
  memory is only quadratically dependent on detuning.
  Moreover, $\tilde{T}^{\text{(m)}}_1$ is limited by the dephasing time of the
  superconducting qubit.
  \textcolor{black}{
  Since a detuned superconducting qubit is
  strongly affected by the environment
 \cite{KakuyanagiMenoSaitoNakanoSembaTakayanagiDeppeShnirman01a,YoshiharaHarrabiNiskanenNakamura01a,
  astafiev2004quantum}, the
 dephasing time of the superconducting qubit becomes as small as tens of
 nanoseconds, which could lead to a severe energy leakage from the memory
  qubits. It is also worth mentioning that, even if we couple a microwave cavity with the memory
  device instead of the superconducting qubit
  \cite{tordrup2008holographic,wesenberg2009quantum,kubo2010strong,schuster2010high,
  abe2011electron}, any imperfection of the cavity 
  in such a coupled system 
  will also cause similar energy leakage from the memory qubits.
  }

  \textcolor{black}{  This kind of acceleration of the energy relaxation
  can be understood in terms of
  the violation of energy conservation caused by the dephasing process,
  which has been discussed for biology systems
  \cite{plenio2008dephasing,rebentrost2009environment,caruso2009highly,fujii2010anti}.
   Also, if we consider the superconducting qubit as a measurement apparatus for the memory qubit, it would be also possible to
  interprete this enhanced relaxation as the appearance of the anti Zeno
  effect induced by false measurements of erroneous apparatus \cite{koshino2004quantum}.
 The decay is accelerated by
 the difference between the energy band of the measurement apparatus and
 the energy of the signal emitted from the unstable state
  \cite{koshino2004quantum}.}
  \textcolor{black}{In our case,
  the detuned superconducting qubit would be interpreted as the erroneous
  apparatus to measure the signal, namely to determine which energy
  etgenstate the memory qubit is in, so that the energy transport from the memory qubit could be accelerated due to the imperfection of the apparatus.
   In addition, a
   similar expression has also
 been used in quantum optics, and is called the scattering
   rate \cite{metcalf1999laser}.
  For example, when we drive the Rabi oscillations of an unstable two-level system with a
  detuned light, the total scattering rate of light from the laser field
  can be also suppressed only quadratically against the detuning
  \cite{metcalf1999laser}. So it
  would be possible to interprete the enhanced
  relaxation rate in our calculation as the scattering rate of the excited
  population, although in our case the scattering is caused by the
  dephasing of the superconducting qubit.}
  
    Regardless of the interpretation of the enhanced relaxation of the memory qubit,
\textcolor{black}{   
our results are of significant importance from a practical
  point of view.
  Since the threshold of the acceptable error rate
for achieving fault tolerant quantum computation is quite small,
typically of the order of
$1\%$ \cite{RHG01a}, it is essential to find ways to store quantum
  states in reliable memory devices isolated from the environment. However, our results show that the standard way to
  decouple the control qubit from the memory qubit by detuning may not
  sufficiently suppress
  the noise in the stored quantum states, which casts a doubt on the
  feasibility of using memory qubit strategies for scalable quantum computation. Therefore, this result
  motivates us to find another decoupling method to protect the memory
  qubits from the noise induced by such anti Zeno relaxation, which we
  discuss later in our paper.}

  It is worth mentioning that this incoherent energy relaxation is much
  more severe than the well known errors caused by the dispersive Hamiltonian.
Without decoherence, the Hamiltonian between the
superconducting qubit and the memory qubit can be represented as a
  dispersive form
$H=\frac{g^2}{\Delta }\hat{\sigma }^{(\text{sc})}_z \hat{\sigma
}^{(\text{m})}_z$ for a large detuning \cite{holland1991quantum}. Therefore, if we tune the
  superconducting qubit so that it is on resonant with a third party system such as
  another qubit for information operations, the superconducting
  qubit in a superposition state induces phase errors in the detuned memory
  qubit. The error rate is
  $\epsilon \simeq \frac{g ^2 t_{\text{I}}}{\Delta }$ where
  $t_{\text{I}}$ is the time required for information operations on
  the third system.
  Fortunately, such information operations can be performed in tens of
  nanoseconds
  and so this kind of phase error can be small. Moreover, as long as the superconducting qubit is
  detuned from any other qubits, the effect of the dispersive Hamiltonian on the memory
  qubit can be negligible
  by polarizing the
  superconducting qubit into the ground state. These results seem to show the suitability of
  this scheme for the long-term storage of information.
  However, as we have shown above, this naive illustration is no longer
  valid when we \textcolor{black}{take into account} the effect of the dephasing from the environment.
  In fact, incoherent energy relaxation
  via the anti Zeno effect occurs whenever the information is stored in
  the memory qubit. In spite of the fact that the memory qubit is
  assumed to retain the information for a long period,
  the information continues to leak during the storage and so the
  total error accumulation will be significant when we try to \textcolor{black}{access}
  the information in the memory after such a storage.

\begin{figure}[h]
       \begin{center}
        \includegraphics[width=7.5 cm]{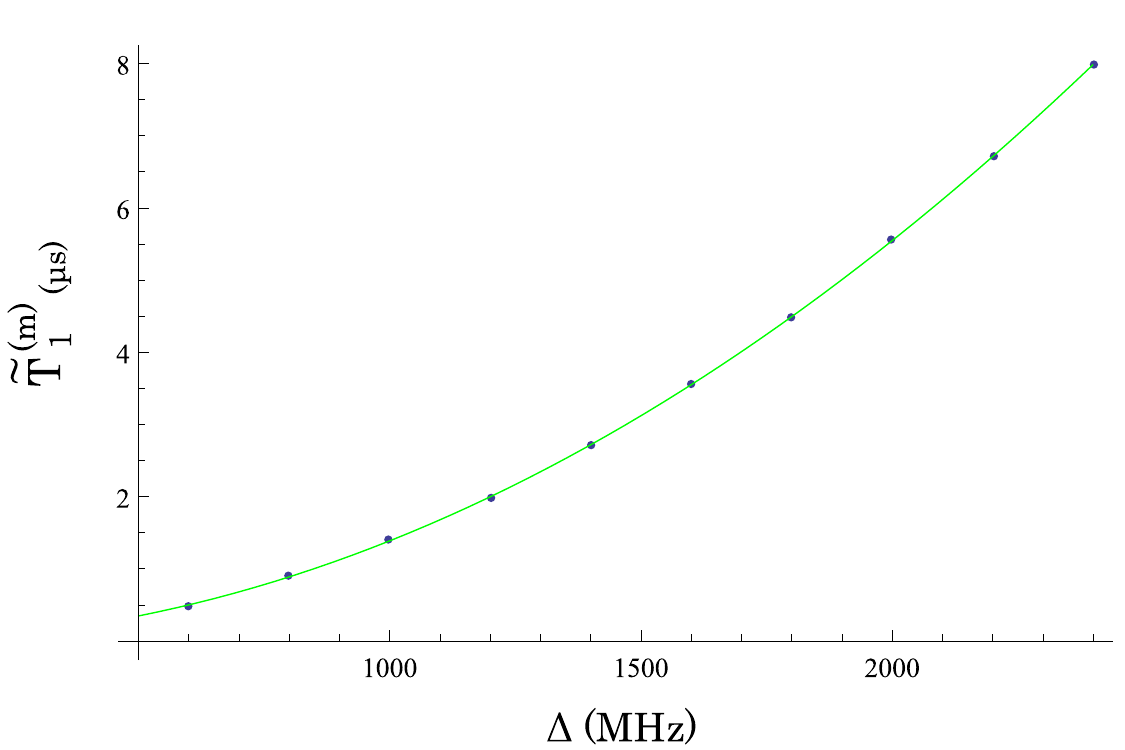} 
        \caption{A plot of the relaxation time of the memory qubit induced by the anti Zeno
        effect of as a function of the detuning. The discrete
        plot denotes a numerical result and the continuous line denotes an
        analytical result. We set parameters as
        $T^{(\text{sc})}_2=10$ ns for the dephasing time of the
        superconducting qubit and at $g/2\pi= 25$ MHz for the coupling strength
        between the qubits.
        These results show that the relaxation time is quadratically
         dependent on
        the energy detuning.
        }\label{compare-graph}
       \end{center}
       \end{figure}

 We have obtained an analytical formula for the effective relaxation time of
 the memory qubit with some approximation.
 It is possible to obtain a more rigorous result by
 solving a Lindblad master equation numerically.
 So we adopt $  \frac{d\rho }{dt}=-i[H,\rho ]-\frac{1}{2T^{(\text{sc})}_2}[\hat{\sigma
   }^{(\text{sc})}_z,[\hat{\sigma }^{(\text{sc})}_z, \rho ]]$
  as the
 master equation \textcolor{black}{where $\rho $ denotes a density matrix for the composed
 system of a
 superconducting qubit and a memory qubit.} We have plotted the effective relaxation time of the
 memory qubit from the numerical solution in Fig. \ref{compare-graph}. The numerical
 solution shows the quadratic dependency of $\tilde{T}^{(\text{m})}_{1}$
 on the detuning, \textcolor{black}{which agrees with the analytical result in Eq. (\ref{effective-relaxation-time}).}
 By fitting the analytical result with this numerical result, we obtain
 $\alpha =0.500$ and we also
 plot the analytical result in Fig. \ref{compare-graph}.
 The behavior of our analytical solution matches the numerical
 solutions, and this shows the validity of our approximation.
\textcolor{black}{ This justifies our interpretation that the dephasing corresponds to 
 non-selective measurements and the energy relaxation in the memory
 qubit is caused by the anti Zeno effect.}
   Surprisingly,
 even for a large detuning such as $\frac{\Delta }{g}\simeq 50$, the
 relaxation time is just a few microseconds. Since a typical
 memory qubit is considered to have a long lifetime of, for example,
 tens of milliseconds
 \cite{tyryshkin2003electron, morton2008solid,morton2007environmental, marcos2010coupling}, this result shows that the actual
 relaxation time induced by imperfect decoupling is much shorter
 than previously expected.

\begin{figure}[h]
       \begin{center}
        \includegraphics[width=8cm]{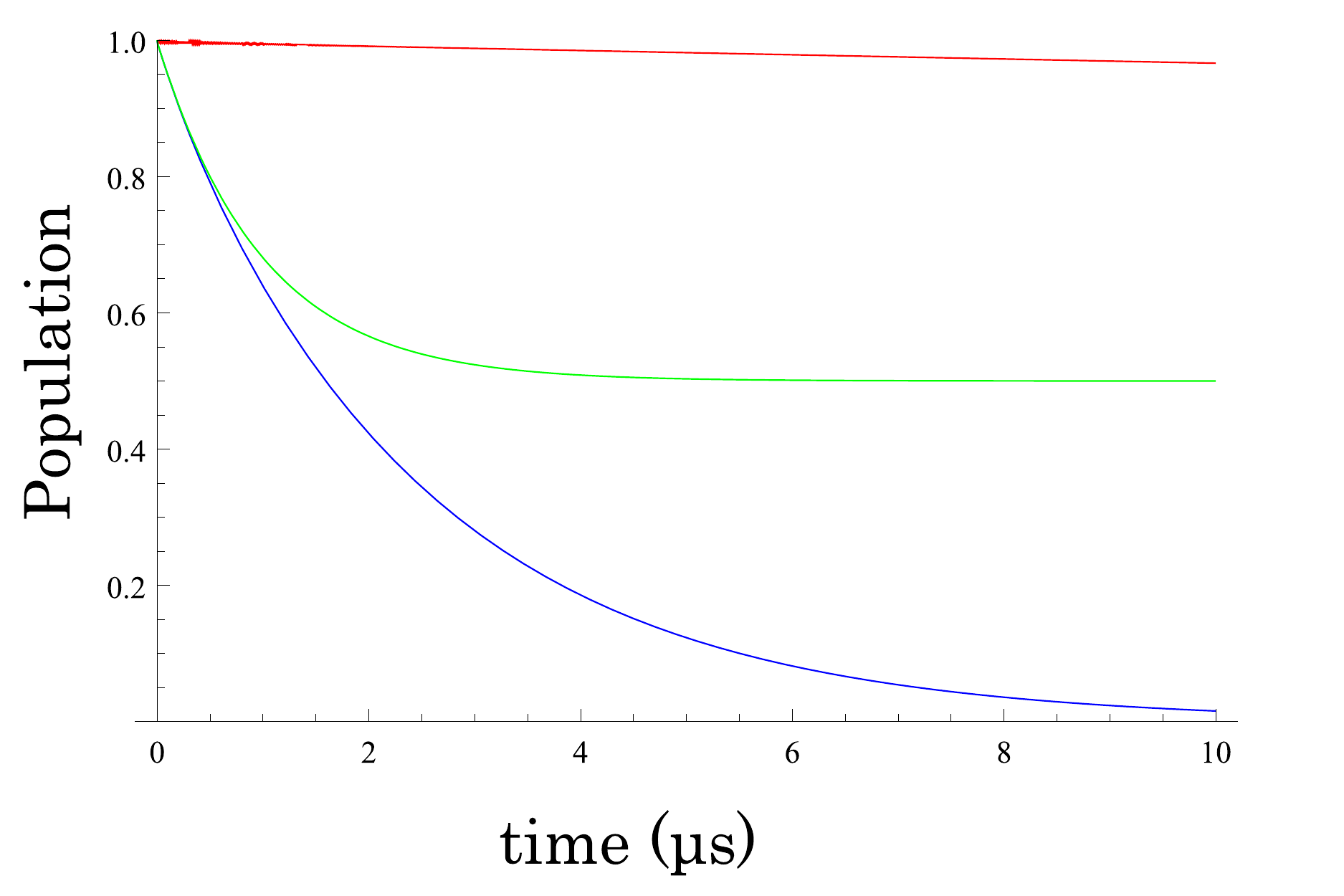} 
        \caption{The population decay of the memory qubit induced by the anti Zeno
        effect as a function of a time. The coupling constant between qubits is set
        at $2\pi \times 25$ MHz.
        From the bottom, we plot a numerical result with $T^{(\text{sc})}_1=400 $ ns and $T^{(\text{sc})}_2=10$
        ns, a numerical result with $T^{(\text{sc})}_2=10$ and
        $T^{(\text{sc})}_1= \infty $,
        and a
        numerical result with $T^{(\text{sc})}_2=\infty $ and $T^{(\text{sc})}_1= 400$ ns.
        }\label{popualtion-graph}
       \end{center}
       \end{figure}

\begin{figure}[h]
       \begin{center}
        \includegraphics[width=8.5cm]{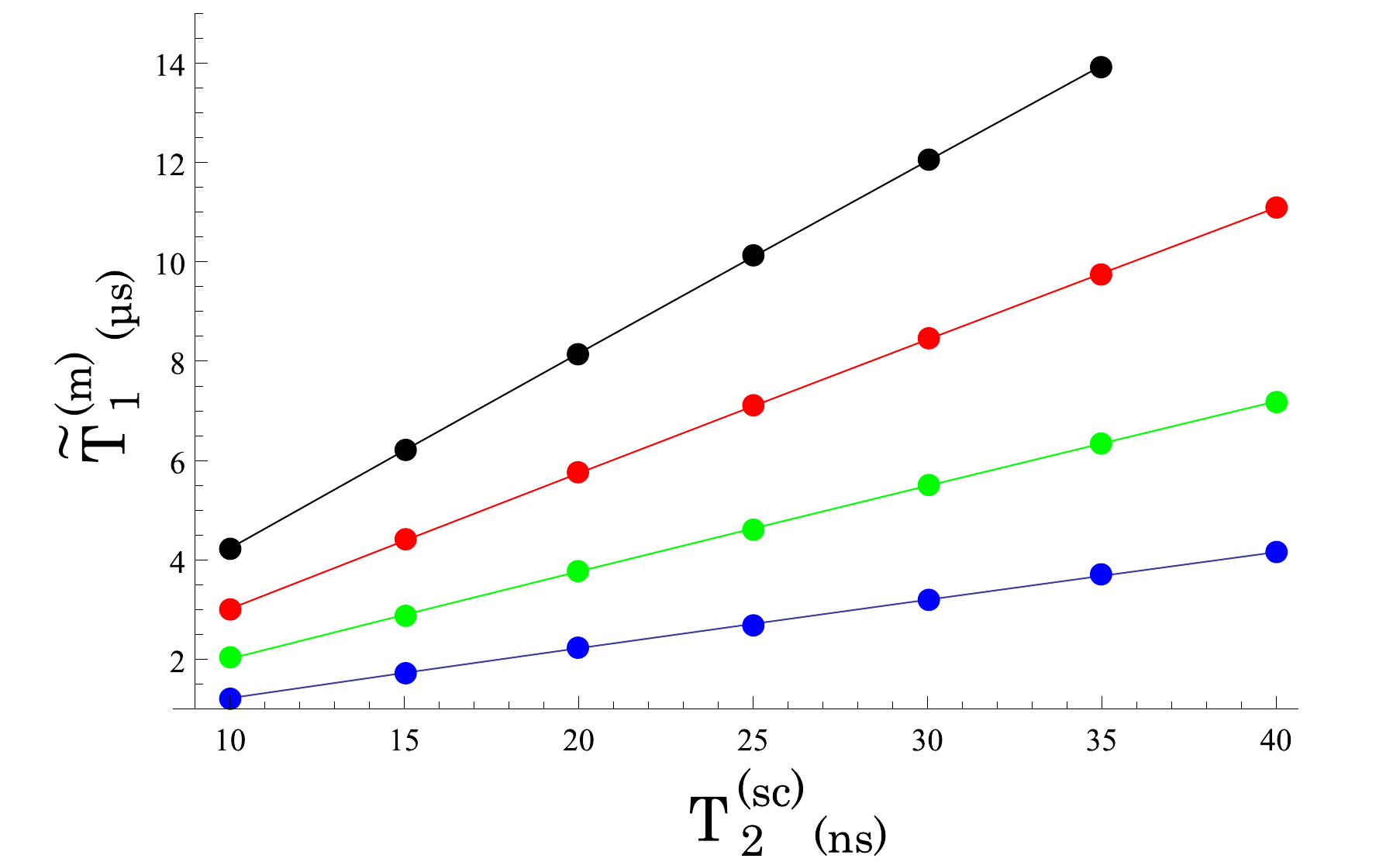} 
        \caption{
        The relaxation time of the memory qubit induced by the anti Zeno
        effect. The horizontal line denotes the dephasing time of the
        superconducting qubit.
        Dots correspond to numerically obtained data.
        Also, continuous lines are drawn through the points as a guide
          to the eye. We set the relaxation time of the superconducting
        qubit at $T^{(\text{sc})}_1=400$ ns and
        the coupling constant between qubits at $g/2\pi= 25$ MHz. The
         lowest line is that where the detuning
        $\Delta /2\pi $ is
        $600$ MHz, and the other lines are where $\Delta /2\pi =800, 1000, 1200$
        MHz respectively.
        }\label{2compare-graph}
       \end{center}
       \end{figure}

 A superconducting qubit can be affected by both dephasing and
 relaxation.  To model both the dephasing and relaxation on the
 superconducting qubit, we add a relaxation term to the Lindblad master
 equation and we adopt the following master equation
 $ \frac{d\rho }{dt}=-i[H,\rho ]-\frac{1}{2T^{(\text{sc})}_2}[\hat{\sigma
   }^{(\text{sc})}_z,[\hat{\sigma }^{(\text{sc})}_z, \rho ]]
  -\frac{1}{2T^{(sc)}_1}(\hat{\sigma }^{(\text{sc})}_+\hat{\sigma
   }^{(\text{sc})}_-\rho +\rho \hat{\sigma }^{(\text{sc})}_+ \hat{\sigma
   }^{(\text{sc})}_- -\hat{\sigma }^{(\text{sc})}_-\rho \hat{\sigma }^{(\text{sc})}_+)$
 where $T^{(\text{sc})}_1$ denotes the relaxation time of a
 superconducting qubit.
 By solving this master equation numerically,
 we are able to plot the population decay behavior of the memory qubit
 \textcolor{black}{as shown in}
 Fig. \ref{popualtion-graph}.
\textcolor{black}{In the absence of the relaxation in the
  superconducting qubit}, the population of the memory qubit decays to
  half of the initial population while the population decays to
  zero under the effect of the relaxation
  on the superconducting qubit. 
  From the numerical solution, we plot
  $\tilde{T}^{\text{(m)}}_1$ as a function of $T^{(\text{sc})}_{2}$ in 
  Fig. \ref{2compare-graph}.
  Even for a long dephasing time and huge
  detuning such as
  $T^{(\text{sc})}_2=35$ ns
  and $\frac{\Delta }{g }=44$, the
  effective relaxation time $\tilde{T}^{\text{(m)}}_1$ is around $14$ $\mu$s, which is much
  shorter than the typical lifetime of the memory qubit \cite{tyryshkin2003electron, morton2008solid,morton2007environmental, marcos2010coupling}.
  Therefore, our results show that the standard strategy to detune the
  superconducting qubit with the memory qubit actually fails due to the energy
  leakage from the memory qubit during the storage of the information.
  \textcolor{black}{It should be noted that, since our model is quite
  general,
     the result here is significant for every
  hybrid system where a device having a short coherence time couples with a memory
 device, as long as the coupling is
 represented by the Jaynes-Cummings model
or the Tavis-Cummings model.}
\section{IV. Overcoming energy relaxation process  by using
 a decoherence free subspace}
  Finally, we discuss possible solutions to the problem of the energy
  relaxation caused by the anti Zeno effect.
  \textcolor{black}{As discussed in the previous section, \textcolor{black}{significant} dephasing of a superconducting
  qubit can nullify the advantage of the
  long lifetime of the memory qubit due to the indirect relaxation, and
  therefore if we are to realize a hybrid system it is crucial that we
  overcome such enhanced relaxation problems.
 As an example, we discuss how to avoid
 this enhanced relaxation when the memory qubit consists of an
 ensemble of microscopic spins with a long life time. 
 }
   When one uses an ensemble composed of $N$
   $\frac{1}{2}$ spins as a memory qubit,
   the excitation of the superconducting qubit is transferred to the
   ensemble and stored as a collective mode. A state with a single
   collective mode in the ensemble is represented as $|W\rangle =\frac{1}{\sqrt{N}}\sum_{l=1}^{L}\hat{\sigma
     }^{(l)}_+|\downarrow \downarrow \cdots \downarrow \rangle $
   where $\hat{\sigma }^{(l)}_+$ denotes the raising operator of a spin
    and $|\downarrow \rangle $ denotes the ground state of a
   single spin. \textcolor{black}{This strategy of utilizing the spin ensemble directly
  coupled with the superconducting qubit as a memory is suggested theoretically
  in \cite{marcos2010coupling}. However, if we adopt their strategy straightforwardly, the
  memory ensemble will suffer from the relaxation induced by the anti
  quantum Zeno effect as we have discussed.}
  So our purpose here is to decouple this excitation in the ensemble from
  the superconducting qubit.
  To achieve this, we can apply a \textcolor{black}{spatial} magnetic field gradient
  $\frac{dB}{dx}$ (T/m) with some
  time duration to the state $|W\rangle $ of the ensemble so that we
  obtain the state $|W _{\theta }\rangle =\frac{1}{\sqrt{N}}\sum_{l=1}^{N}e^{i\theta l}
    \hat{\sigma
    }^{(l)}_+|\downarrow \downarrow \cdots \downarrow \rangle $.
  Here, we have $\theta =\tau  \mu \frac{dB}{dx} \Delta x $ where $\mu $
  denotes the magnetic moment of the spin, $\tau $ denotes application
  time of such a field gradient, and $\Delta x$ denotes the distance
  between the spins.
   Since we have $\langle W|W_{\theta }\rangle
  =\frac{1}{N}\sum_{l=1}^{N}e^{i\theta l}$, the state $|W_{\theta
  }\rangle $ becomes orthogonal with the state $|W\rangle $ for $\theta
  N=2\pi $, and therefore we can decouple the ensemble from the superconducting
  qubit. For reasonable parameters such as
  $2\pi \times 28$ GHz/T for the Zeeman splitting 
  and $N\Delta x=20 $ $\mu $m for the ensemble length, we need a field
  gradient $10$ T/m to achieve the orthogonal state in
  hundreds of nanoseconds.   \textcolor{black}{
  This idea of applying a field gradient was developed in the field of holographic quantum
  computation for different purposes
  \cite{tordrup2008holographic,wesenberg2009quantum}.
  In holographic quantum
  computation, we utilize a hybrid system composed of a
  superconducting qubit, a spin ensemble, and a microwave cavity. 
  A field gradient will be applied to transfer the collective excitation
  of the ensemble to another spin wave mode so that
  we can store many qubits in one ensemble. However, the effect of
  relaxation enhanced by the anti Zeno effect, in other words, indirect
  relaxation caused by the imperfect decoupling, has not been discussed in
  previous research.
  We use
  the technique to apply a field gradient for the efficient decoupling
  of the superconducting qubit from the spin ensemble so that a long
  time storage of quantum states should be possible in the memory.} \textcolor{black}{
Although we have mentioned a solution for the spin ensemble memory here, the basic
  idea could be applied to many other systems. }\textcolor{black}{We discuss this
  point briefly, although the details will
 be published elsewhere.}
\textcolor{black}{As the quantum memory, one can use multiple
  memory qubits collectively coupled with the control
superconducting qubit. It becomes possible
to make a decoherence free subspace \cite{lidar21998decoherencefreesubapace}
against the interaction from the control qubit
  as follows:
  The Hamiltonian between the control superconducting qubit and memory
  qubits would be $H=H_S^{(\text{sc})}+H_S^{(\text{m1})}+H_S^{(\text{m2})}
  +H_I^{(\text{sc},\text{m1})}+H_I^{(\text{sc},\text{m2})}$
where we have $H^{(j)}_S=\frac{\omega _j}{2} \hat{\sigma }_z ^{(j)}$and $H_I^{(j,k)}=
g(\hat{\sigma }_+^{(j)}\hat{\sigma }_-^{(k)}+
   \hat{\sigma }_-^{(j)}\hat{\sigma }_+^{(k)} )$. An excitation of the
  control superconducting qubit can be transferred into the collective excitation on the
  memory qubits such as $\frac{1}{\sqrt{2}}(|01\rangle _{\text{m1,m2}}+|10\rangle _{\text{m1,m2}})$.
By performing a local phase flip on one of the memory qubits, we have $\frac{1}{\sqrt{2}}(|01\rangle _{\text{m1,m2}}-|10\rangle _{\text{m1,m2}})$,}
\textcolor{black}{
 which is completely orthogonal to the previous state and decoupled from
 the controllable qubit.
 In other words, this state is in the decoherence free subspace for the
 indirect relaxation.
Therefore, the memory qubit is immune to the energy leakage induced by the quantum
anti Zeno effect.}
\textcolor{black}{
So this decoupling using a decoherence free subspace is much more robust
than the traditional way of decoupling the system by detuning.
Recently, the coherent control of three qubits and 
the generation of an entanglement between them have
been demonstrated by using superconducting qubits
\cite{altomare2010tripartite,neeley2010generationetal,dicarlo2010preparation},
and therefore our suggestion here is feasible even with current technology.}
\section{V. Conclusion}
In conclusion, we have studied the indirect energy relaxation in a hybrid system consisting
of a superconducting qubit and a memory qubit. \textcolor{black}{
Even if we employ the detuning
between a memory qubit and a superconducting qubit to decouple them,
the dephasing of the superconducting qubit could significantly affect the coherence of the memory qubit.
The violation of the
energy conservation law due to dephasing of the superconducting qubit
  induces an incoherent energy leakage from the memory, which leads to a significant degradation in the lifetime of the quantum
  memory. This can be interpreted as a manifestation of the quantum anti
  Zeno effect.
  If this indirect relaxation is inevitable, a hybrid scheme
  with a superconducting qubit would
  not be promising unless we can succeed in making the coherence time of the
  superconducting qubit longer than the present value.
  However, we can find a
  possible solution to this problem, for example, via  decoupling from the
  superconducting qubit using a decoherence free subspace for the
  memory qubits.}
  Our model is quite general, and
  therefore the results reported here can be
  applied to many systems. 
We thank K Semba, B. Munro, X. Zhu, K. Fujii, S. Saito, H. Tanji, and K. Kakuyanagi for useful discussions.
This work is partially supported by KAKENHI (Grant-in-Aid for Scientific
Research A 22241025).


\end{document}